\documentclass[a4paper,11pt]{article}
\pdfoutput=1 

\usepackage{jinstpub} 

\usepackage{lineno}

\title{\boldmath CMS Muon Drift Tubes HL-LHC Slice Test}

\author{B\'arbara \'Alvarez Gonz\'alez on behalf of the CMS Collaboration}

\affiliation{Universidad de Oviedo, Instituto Universitario de Ciencias y Tecnolog\'ias Espaciales de Asturias (ICTEA), Spain}
\emailAdd{balvarez@cern.ch}

\abstract{The electronics of the CMS Drift Tube (DT) chambers need to be replaced during Long Shutdown 3 to tolerate the High Luminosity LHC (HL-LHC) data taking conditions. The first DT prototypes of the HL-LHC electronics, the on-detector board (OBDT), have been installed in the DT chambers of one sector, out of 60 sectors, and integrated into the central data acquisition and trigger systems. The signals from the chambers are split and reach both the legacy and Phase-2 demonstrator chains, which will allow them to operate in parallel during LHC collisions.}


\proceeding{TWEPP 2021 Topical Workshop on Electronics for Particle Physics\\
  20-24 September 2021\\
  Virtual}

\begin{document}
\maketitle
\flushbottom

\section{Introduction}
\label{sec:intro}
After delivering a total integrated luminosity of more than 160~fb$^{-1}$ at the end of Run 2, at the beginning of 2019, the Large Hadron Collider (LHC) was shut down for two years (Long Shutdown 2, LS2) to upgrade the accelerator-chain and detectors upgraded for the High Luminosity LHC (HL-LHC) phase. 
The Compact Muon Solenoid (CMS)~\cite{cms} Drift Tube (DT) muon detector system, originally built for processing the instantaneous and integrated luminosities expected for the LHC, will now be required to process the HL-LHC data, which has a factor five larger instantaneous luminosity.  Consequently, the radiation levels are expected to exceed the LHC integrated luminosity by a factor of ten. During this LS2, CMS aims to upgrade its electronics and detector performance to improve the data taking and to ensure a precise reconstruction of all the particles in the high pile-up conditions of HL-LHC using the existing DT chambers. 

\section{CMS Drift Tubes}
\label{sec:dts}
The Drift Tube chambers are one of the important parts of the CMS muon system responsible for identifying, measuring and triggering on muons by the precise measurement of their position and momentum. The electronics of the CMS DT chambers will need to be replaced for the HL-LHC operation~\cite{hllhc}.
\begin{figure}[htbp]
	\centering 
	\includegraphics[width=.33\textwidth]{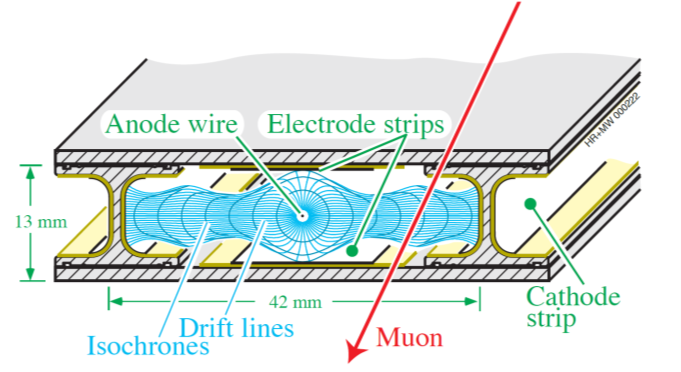}
	\includegraphics[width=.3\textwidth]{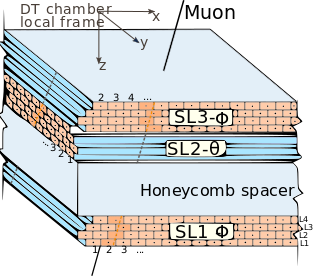}
	\includegraphics[width=.33\textwidth]{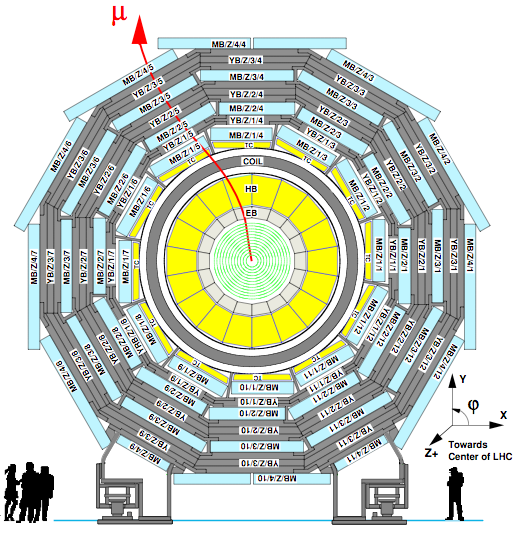}

	\caption{\label{fig:i} DT cell (left), DT chamber (middle), CMS tranverse cut (right).}
\end{figure}
 
The basic element of the DT system is the DT drift cell, shown in Fig.~\ref{fig:i} on the left side, a rectangular cell of 42 x 13~mm$^{2}$, with a gas mixture of 85\% Ar and 15\% CO$_{2}$ and resulting in a roughly constant drift velocity of 54 $\mu$m/ns. There are approximately 172000 DT drift cells. A DT chamber, shown in Fig.~\ref{fig:i} in the middle part, is made of parallel layers of cells grouped in four layers that form a super layer (SL); two SLs in r-$\phi$ and one SL r-z (for the three innermost stations). A DT muon station consists of an assembly of chambers located at a fixed value of radial distance R, with four barrel stations labelled as MB1 to MB4, where MB1 is the closest station to the interaction point. Along the beam axis z DTs are divided into five slices, called wheels (see Fig.~\ref{fig:i} for one wheel), with wheel 0 centered at z=0 and wheel+1 and wheel+2 in the positive z direction and wheel-1 and wheel-2 in the negative z direction. Within each wheel, chambers are arranged in twelve azimuthal sectors.

\section{Slice Test}
\label{sec:obdt}
For a slice test of the HL-LHC electronics, prototypes of the on-detector board for the drift tube chambers (OBDTs) have been installed in a single sector (wheel+2, sector 12) and integrated into the central data acquisition and trigger systems during LS2. The four chambers in this sector were instrumented with OBDT prototypes~\cite{obdt}. The DT chamber front-end pulses carrying the time information of the chamber hits are sent to both the existing on-detector electronics (minicrate) and through the OBDTs via specifically designed splitter boards that preserve the signal integrity. Thirteen OBDTs distributed in five mechanical frames which also take care of the thermal interface to the water cooling loop are installed in this sector. The Phase-2 back-end functionality (slow control, trigger generation and DAQ) is implemented in firmware running on DT uTCA boards (TM7~\cite{track}) developed for the Phase-1 upgrade. 
The OBDT boards use a Polarfire FPGA and they digitize the pulses that come from the front-end boards located inside the chambers. The FPGAs send digitalized and formatted data to five AB7 boards through its five optical links. A new trigger system based on high performing FPGAs is being designed that will be capable of providing precise muon reconstruction and bunch crossing identification.

\begin{figure}[htbp]
	\centering 
	\includegraphics[width=.48\textwidth]{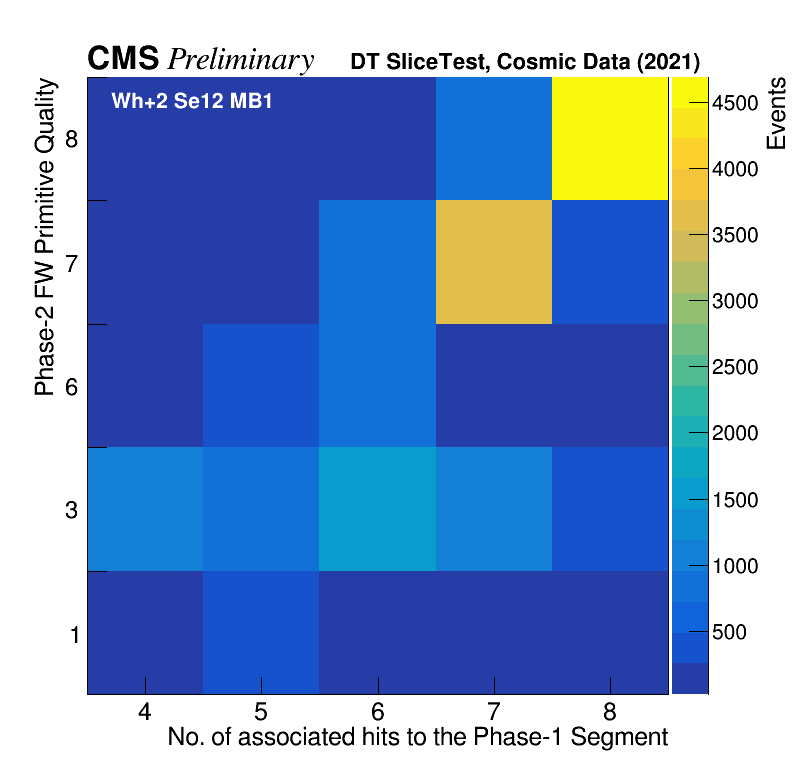}	
	\includegraphics[width=.48\textwidth]{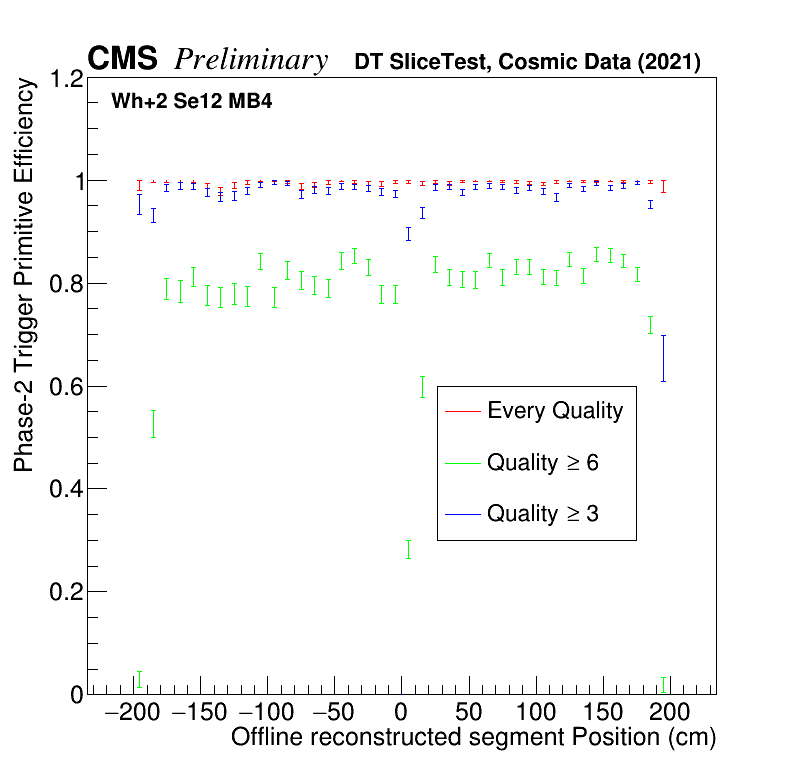}
	\caption{\label{fig:2Dandeff} Left: 2D distribution of the TP Quality vs the number of hits associated to the offline reconstructed segment. Right: Efficiency of finding a Phase-2 TP in any BX with respect to the local position of the offline segment reconstructed out of
		hits detected by the Phase-1 system.}
\end{figure}

The slice test results presented here were obtained in 2021 using cosmic rays. As mentioned previously, the signals from the chambers are split and reach both the legacy existing Phase-1 electronics and Phase-2 demonstrator chains, which will allow them to operate in parallel during LHC collisions. Figure~\ref{fig:2Dandeff} shows the 2D distribution of the Phase-2 Trigger Primitive (TP) Quality obtained by the AB7 board as a function of the number of hits associated to the offline reconstructed segment (phi view). The quality criteria is explained in Table~\ref{tab:qual}. 
\begin{center}
	\begin{table}[ht]
		\centering
	\caption{Trigger Primitive Quality criteria}
	\begin{tabular}{c|c}
		Quality & Description  \\\hline
		1& 3 hit track    \\
		3& 4 hit track   \\
		6& 3+3 hit track   \\
		7& 3+4 hit track   \\
		8& 4+4 hit track   \\
	\end{tabular}
	\label{tab:qual}
	\end{table}
\end{center}
The highest number of events corresponds to the highest quality, which matches four plus four hits from a Phase-2 TP with eight hits from an offline reconstructed segment. The efficiency of finding a Phase-2 TP in any bunch crossing (BX) with respect to the local position of the offline segment reconstructed out of hits detected by the Phase-1 system is shown in Fig.~\ref{fig:2Dandeff}. It includes every primitive (red), primitives built with more than 4 hits (blue), and primitives with more than 6 hits, i.e. 3 or more hits per superlayer (green) in MB4. Selected segments are built with more than 4 hits and have an inclination in the radial coordinate smaller than 30 degrees with respect to the direction perpendicular to the chamber. No geometrical matching between the offline segment and the TP is required.

\begin{figure}[htbp]
	\centering 
	\includegraphics[width=.5\textwidth]{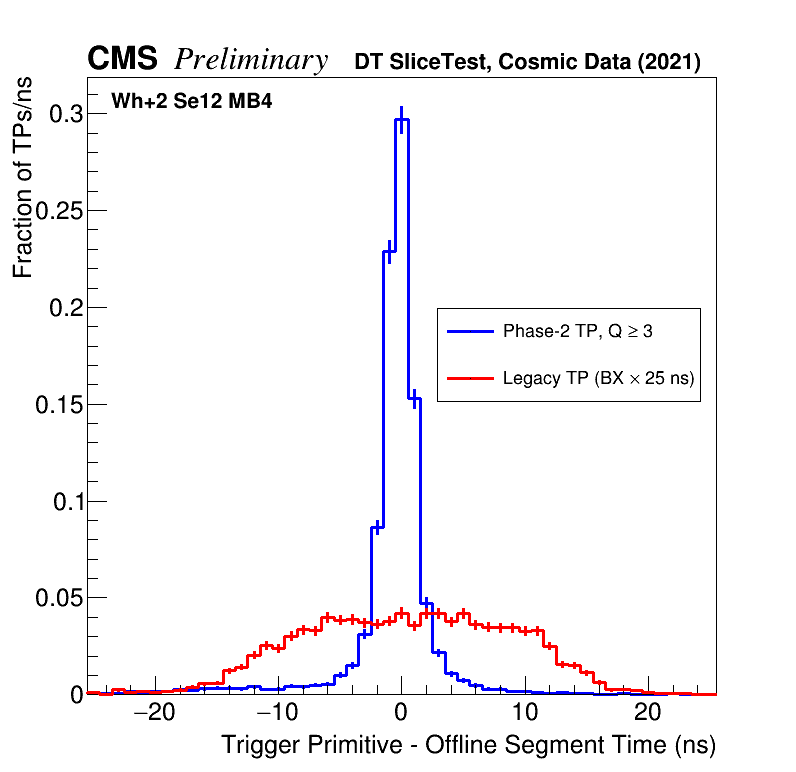}
	\caption{\label{fig:DeltaTime} Difference between trigger primitive’s time and the offline reconstructed segment time in a cosmic sample.}
\end{figure}

The time difference between trigger primitive time and the offline reconstructed segment time in a cosmic sample is shown Fig.~\ref{fig:DeltaTime}. For Phase-2 only primitives fitting at least 4 hits are compared with the legacy system. The core resolution of the Phase-2 distribution is a few ns, whereas for the Phase-1 system the trigger output time is given in bunch crossing units (25~ns step). The improved online time resolution in Phase-2 reflects in this particular sample (unbunched cosmic muons) as a lower fraction of triggers at a wrong bunch crossing, i.e. 12.5~ns away from the time the muon crossed the chamber. 

\section{Summary}
\label{sec:summary}
For HL-LHC the DT electronics will be fully replaced while
keeping the existing chambers, the prototypes of the Phase-2 On Board DT electronics are integrated on site in CMS as part of the DT Slice Test. The full Slice Test data taking chain has been operated very satisfactorily, showing the optimal efficiency of the designed Phase-2 electronics and good performance is obtained. The hit detection and offline reconstruction is are comparable to the Phase-1 system, already exploiting the ultimate DT cell resolution. Moreover, a significant improvement in the Phase-2 Level 1 DT local trigger resolution is also reached. We plan to run this Phase-2 parallel system in collisions during Run 3, which will allow us to test final preproduction prototypes under realistic conditions (radiation, magnetic field) and further refine the trigger algorithms.

\end{document}